\documentstyle[12pt]{article}

\newlength{\extraspace}
\newlength{\extraspaces}
\catcode`\@=11

\def\numberbysection{\@addtoreset{equation}{section}
\def\theequation{\arabic{section}.\arabic{equation}}}

\newcommand{\be}{\begin{equation}
\addtolength{\abovedisplayskip}{\extraspaces}
\addtolength{\belowdisplayskip}{\extraspaces}
\addtolength{\abovedisplayshortskip}{\extraspace}
\addtolength{\belowdisplayshortskip}{\extraspace}}
\newcommand{\ee}{\end{equation}}
\newcommand{\ba}{\begin{eqnarray}
\addtolength{\abovedisplayskip}{\extraspaces}
\addtolength{\belowdisplayskip}{\extraspaces}
\addtolength{\abovedisplayshortskip}{\extraspace}
\addtolength{\belowdisplayshortskip}{\extraspace}}
\newcommand{\ea}{\end{eqnarray}}
\newcommand{\nonu}{\nonumber \\[.5mm]}
\newcommand{\tr}{\, {\rm tr} \,}

\newcommand{\D}{{\cal D}}

\newcommand{\VEV}[1]{\left\langle {#1} \right\rangle}

\newcommand{\R}{\bf R}

\newcommand{\tg}{\tilde{g}}

\newcommand{\lt}{\left(}
\newcommand{\rt}{\right)}

\newcommand{\la}{\langle}
\newcommand{\ra}{\rangle}
\newcommand{\su}{\sum^{N}_{i=1}}
\newcommand{\oon}{\frac{1}{N}}

\newcommand{\pa}{\partial}
\newcommand{\pr}{\prod^N_{i=1}}
\renewcommand{\l}{\lambda}
\renewcommand{\D}{\Delta}
\newcommand{\ntr}{\oon \tr}

\newcommand{\ddn}{\frac{\pa}{\pa N}}

\begin{document}
\begin{titlepage}
\addtolength{\baselineskip}{.7mm}
\thispagestyle{empty}
\begin{flushright}
TIT/HEP--227 \\
NUP-A-93-12 \\
hepth@xxx/9307065\\
June, 1993
\end{flushright}
\vspace{2mm}
\begin{center}
{\large{\bf
Renormalization Group Approach to Discretized Gravity
\footnote{An updated version of the talk presented
by N.~Sakai
at the First Pacific Winter School for Theoretical Physics,
Sorak Mountain Resort, Korea, (22--26, February 1993), to appear in
the Proceedings.}
}} \\[8mm]
%
{\sc Saburo Higuchi}
\footnote{{\tt e-mail: hig@phys.titech.ac.jp}, JSPS fellow}\\[2mm]
{\it Department of Physics, Tokyo Institute of Technology, \\[1mm]
Oh-okayama, Meguro, Tokyo 152, Japan} \\[3mm]
{\sc Chigak Itoi}
\footnote{\tt e-mail: itoi@phys.titech.ac.jp} \\[2mm]
{\it Department of Physics and Atomic Energy Research Institute, \\[1mm]
College of Science and Technology, Nihon University, \\[1mm]
Kanda Surugadai, Chiyoda, Tokyo 101, Japan} \\[3mm]
{\sc Shinsuke Nishigaki}
\footnote{\tt e-mail: nsgk@phys.titech.ac.jp},
and
{\sc Norisuke Sakai}
\footnote{\tt e-mail: nsakai@phys.titech.ac.jp} \\[2mm]
{\it Department of Physics,
Tokyo Institute of Technology, \\[1mm]
Oh-okayama, Meguro, Tokyo 152, Japan} \\[6mm]
{\bf Abstract}\\[4mm]
\parbox{13cm}{\hspace{5mm}
We summarize our renormalization group approach for the
vector model as well as the matrix model which are
the discretized quantum gravity in one- and
two-dimensional spacetime.
A difference equation is obtained which relates
free energies for neighboring values of $N$.
The reparametrization freedom in field space is
formulated by means of the loop equation.
The reparametrization identities reduce the flow in the infinite
dimensional coupling constant space to that in
finite dimensions.
The matrix model gives a nonlinear differential equation
as an effective renormalization group equation.
The fixed point and the susceptibility exponents can be determined
even for the matrix models in spite of the nonlinearity.
They agree with the exact result.
}
\end{center}
\end{titlepage}
\begin{center}
{\large{\bf
Renormalization Group Approach to Discretized Gravity
}} \\[8mm]
%
{\sc Saburo Higuchi}
\\[2mm]
{\it Department of Physics, Tokyo Institute of Technology, \\[1mm]
Oh-okayama, Meguro, Tokyo 152, Japan} \\[3mm]
{\sc Chigak Itoi}
\\[2mm]
{\it Department of Physics and Atomic Energy Research Institute, \\[1mm]
College of Science and Technology, Nihon University, \\[1mm]
Kanda Surugadai, Chiyoda, Tokyo 101, Japan} \\[3mm]
{\sc Shinsuke Nishigaki}
and
{\sc Norisuke Sakai}
\footnote{The speaker.}
\\[2mm]
{\it Department of Physics,
Tokyo Institute of Technology, \\[1mm]
Oh-okayama, Meguro, Tokyo 152, Japan} \\[6mm]
{\bf Abstract}\\[4mm]
{\parbox{13cm}{\hspace{5mm}
We summarize our renormalization group approach for the
vector model as well as the matrix model which are
the discretized quantum gravity in one- and
two-dimensional spacetime.
A difference equation is obtained which relates
free energies for neighboring values of $N$.
The reparametrization freedom in field space is
formulated by means of the loop equation.
The reparametrization identities reduce the flow in the infinite
dimensional coupling constant space to that in
finite dimensions.
The matrix model gives a nonlinear differential equation
as an effective renormalization group equation.
The fixed point and the susceptibility exponents can be determined
even for the matrix models in spite of the nonlinearity.
They agree with the exact result.
}}
\end{center}
\setcounter{section}{0}
\setcounter{equation}{0}
%
%
\section {Introduction}
The long-standing challenge for the quantum theory of gravity
has met at least a partial success in low dimensions.
The two-dimensional quantum gravity is particularly interesting
from the viewpoint of string theories too.
The discretized approach such as matrix models
offers a possibility for a nonperturbative treatment
\cite{BIPZ}--\cite{GIMO}.
Exact solutions of the matrix model
have been obtained for two-dimensional quantum gravity coupled to
minimal conformal matter with central charge $c \le 1$.
It has been very difficult to obtain results
 for two-dimensional quantum gravity coupled to
conformal matter with central charge $c > 1$.
Although one can easily write down matrix
models for cases with $c > 1$, these models are not solvable up
to now\cite{KAZ}.
The numerical simulations suggest that it is not at all
obvious if a matrix model
candidate to describe a $ c>1 $ model has the
continuum description \cite{DFJ}.
Therefore it is useful to obtain
approximation schemes which
enable us to calculate critical coupling constants and
critical exponents for unsolved matrix models, especially for $c>1$.
In order to make use of such a scheme, we first need to make sure
that the approximation method gives correct results for the
exactly solved cases.
\par
Br\'ezin and Zinn-Justin have proposed a renormalization group
approach to the matrix model\cite{BRZJ}.
Consequences of their approach have been examined by several groups
\cite{ALDA}.
A similar approach has been advocated previously for
the $1/N$ expansion in somewhat different contexts \cite{CARL}.
The vector model has been proposed for a discretized
one-dimensional quantum gravity, in the same way as
the matrix model for a discretized
two-dimensional quantum gravity\cite{VM0}.
Recently we have analyzed the vector model by means of the
renormalization group approach and
have clarified its validity and meaning \cite{HIS}.
More recently, we have succeeded in extending our analysis of
the renormalization group approach to matrix models \cite{HINS}.
\par
The purpose of this paper is to present a new simplified derivation
of our results on vector
models and to summarize some of our results on matrix models in
ref.\cite{HINS}.
For the vector model, a difference equation is obtained relating
the free energy $-\log Z_{N-2}(g)$
to the free energy $-\log Z_N(g - 2\delta g)$ with
the coupling constant shifts $\delta g_k$
of order $1/N$ in infinitely many
coupling constants.
We obtain infinitely many identities which
express the freedom to reparametrize the field space.
Thanks to these identities, we can rewrite the flow
in the infinite dimensional coupling constant space
as an effective flow in the space of finite
number of coupling constants.
In this note, we shall simplify the derivation of the reparametrization
identities by using a new method similar to the loop equation
for the matrix model.
The resulting effective beta function determines the fixed
points and the susceptibility exponents which agree with
the exact results \cite{HIS}.

We also find that a similar procedure works for
matrix models.
Namely we can obtain a difference equation relating the free
energy
$-\log Z_{N+1}(g)$ to the free energy $-\log Z_N(g + \delta g)$ with
shifts $\delta g$ of order $1/N$ in infinitely many coupling constants.
We find infinitely many identities expressing the reparametrization
freedom of matrix variables. The identities are sometimes called
loop equations \cite{GIMO}, \cite{FKN} and enable us
to rewrite the flow
as an effective renormalization group flow in the space of finite
number of coupling constants.
A new characteristic feature of the matrix model is
that the resulting effective renormalization group equations
is nonlinear contrary to the vector model.
In spite of the nonlinearity, we can obtain the fixed points and
the critical exponents \cite{HINS}.
As an explicit example,
we analyzed the one matrix model with a single coupling constant and
find a complete agreement with the exact results for the
first critical point $m=2$ \cite{HINS}.
\par
%
%
\section {Renormalization Group Approach for Vector Models}
It has been proposed that the free energy $F(N,g)$ of the matrix model
should satisfy the following renormalization group equation
 \cite{BRZJ}
\be
\biggl[N{\partial \over \partial N}
-\beta (g) {\partial \over \partial g}
+\gamma(g)\biggr] F (N, g)=r(g),
\label{eqn:matrixrge}
\ee
where $\beta(g)$ is called the beta function.
The anomalous dimension and the inhomogeneous term are
denoted as $\gamma(g)$ and $r(g)$ respectively.
A fixed point $g_*$ is given by a zero of the beta function.
In the following we shall show that this renormalization group equation
is valid for vector models.
\par
The partition function of the $O(N)$ symmetric vector model is given by
\begin{equation}
  Z_N(g) = \int d^N \phi
  \exp \left[ -N \sum_{k=1}^{\infty} \frac{g_k}{2k} (\phi^2)^k\right],
  \label{eqn:vector_action}
\end{equation}
where $\phi$ is an $N$ dimensional real vector
\cite{VM0}.
Here we introduce infinitely many coupling constants $g_k$,
since we need all possible induced interactions after a
renormalization group transformation even if we start with a few
coupling constants only.
The $1/N$ expansion of the logarithm of the partition function
gives contributions from $h$ loops as terms with $N^{1-h}$.
The vector model  has the double scaling limit
$N \rightarrow \infty $ with $N^{1/\gamma_1}(g-g_\ast)$ fixed,
where
the singular part of the free energy satisfies the scaling law
\cite{VM0}
\be
 -\log \left.\biggl[{Z_N(g) \over Z_N(g_1\not=0,g_k=0 \; (k\geq2))}
 \biggr]\right|_{sing}
= \sum_{h=0}^{\infty} N^{1-h}
(g - g_*)^{2 - \gamma_0-\gamma_1 h} a_h + \cdots.
\ee
By inserting this form, one find that the critical point
corresponds to the fixed point $\beta(g_*)=0$ and that
susceptibility exponents $\gamma_0, \gamma_1$ are related to the
derivative of the beta function
\be
\gamma_1 = 1 / \beta'(g_*), \qquad
\gamma_0 = 2- \gamma(g_*) / \beta'(g_*).
 \label{eqn:vectorgamma_and_beta}
\ee

In order to obtain a difference equation,
we start with the partition
function $Z_{N-2}(g)$.
After integrating over angular coordinates in ${\R}^{N-2}$,
we perform a partial integration in the radial coordinate
$x=\phi^2$
\begin{eqnarray}
  Z_{N-2}(g)
    &\!\!\! = &\!\!\! \frac{\pi^{\frac{N}{2}-1}}{\Gamma(\frac{N}{2}-1)}
          \int_{0}^{\infty} dx\;
       x^{N/2-2} \exp\left[-(N-2)\sum_{k=1}^{\infty}
          \frac{g_k}{2k}x^k\right] \nonu
    &\!\!\! = &\!\!\! \frac{\pi^{\frac{N}{2}-1}}{\Gamma(\frac{N}{2}-1)}
          \int_{0}^{\infty} dx\;
       x^{N/2-1} \left(\sum_{k=1}^{\infty} g_k x^{k-1}\right)
         \exp \left[-(N-2)\sum_{k=1}^{\infty} \frac{g_k}{2k}x^k\right].
   \label{eqn:radial_int}
\end{eqnarray}
Identifying the  right hand side with
$((N-2) g_1/ 2\pi) Z_{N}(g - 2 \delta g)$,
we obtain a difference equation for the
logarithm of the partition function
\begin{eqnarray}
\lefteqn{[(-\log Z_{N}(g))-(-\log Z_{N-2}(g))]
- \log\frac{(N-2) g_1}{2\pi}}
 \nonumber \\
 &\!\!\! =&\!\!\! - [ ( -\log Z_{N} ( g - 2 \delta g))
 - ( -\log Z_{N}(g))].
  \label{eqn:difference_equation}
\end{eqnarray}
where the shifts $\delta g_k$ of the coupling constants are found to be
\begin{equation}
  \sum_{k=1}^{\infty} \frac{g_k}{k} x^k
  + \log \left(\sum_{k=1}^{\infty} \frac{g_k}{g_1} x^{k-1}\right)
  =
  N \sum_{k=1}^{\infty} \frac{\delta g_k}{k}    x^k.
   \label{eqn:shift}
\end{equation}
We stress
that no approximation is employed to obtain this difference
eq. (\ref{eqn:difference_equation}) apart from neglecting the
higher order terms in $1/N$ contrary to the perturbative method
advocated in ref.\cite{BRZJ}.
\par
One can bring the quadratic
term in the potential to the standard form $\phi^2/2$
since $g_1$ can be absorbed
by a rescaling $g_1 \phi^2 \rightarrow \phi^2$. We have
\begin{equation}
 Z_N(g_1,g_2,g_3,\ldots) = g_1^{-N/2} Z_N(1, \tg_2,\tg_3,\ldots),
\qquad
 \tg_k = g_k / g_1^k.
\label{rescalingid}
\end{equation}
We shall define the free energy for the vector model
\begin{equation}
F(N,g_1,g_2,g_3,\cdots) =
- \frac{1}{N} \log {Z_N(g_1,g_2,\cdots) \over Z_N(g_1,0,\cdots)}
 =
F(N,\tg_2,\tg_3,\cdots).
\label{eqn:freeenergy}
\end{equation}
We sometimes write the free energy as a function of
$\tg_2, \tg_3, \cdots$, since it is independent of $g_1$
because of the rescaling identity (\ref{rescalingid})
if one uses $g_1, \tg_2, \cdots $ as independent variables.
We denote the partial derivatives with respect to
$g_1, \tilde g_2, \tilde g_3, \cdots$ by $|_{\tilde g}$, and
those with $g_1, g_2, g_3, \cdots$
by $|_{g} $
\begin{equation}
 \left.\frac{\partial}{\partial g_1}\right|_g
=\left.\frac{\partial}{\partial g_1}\right|_{\tg}
   - \sum_{k=2}^{\infty} k \frac{\tg_k}{g_1}
               \left.   \frac{\partial}{\partial    \tg_k}\right|_{\tg},
\qquad
 \left.\frac{\partial}{\partial g_k}\right|_g
= \frac{1}{g_1^k} \left. \frac{\partial}{\partial \tg_k}\right|_{\tg} .
\end{equation}

In the $N \rightarrow \infty$ limit, we can obtain a
differential equation from the
difference equation (\ref{eqn:difference_equation})
as a renormalization group equation for the free
 energy $F$
\ba
\left[N \frac{\partial}{\partial N}
 +1   \right]F(N,g)
&\!\!\!=&\!\!\!     - \frac{1}{2} + \sum_{k=1}^{\infty}
N \delta g_k \left.   \frac{\partial F(N,g)}{\partial g_k} \right|_{g}
\equiv
G, \nonu
G
&\!\!\!=&\!\!\!
      - \frac{1}{2}
+ N \delta g_1 \frac{1}{2g_1} +
  \sum_{k=2}^{\infty}
N \left( \frac{\delta g_k}{g_1^k} - \frac{\delta g_1}{g_1} k \tg_k\right)
\left.     \frac{\partial F(N,\tg)}{\partial \tg_k} \right|_{\tg}
{}.
 \label{eqn:renorm_all_coupling}
\ea
This renormalization group equation clearly shows a flow in the
infinite dimensional coupling constant space as is usual in the
Wilson's renormalization group approach.
Eq.(\ref{eqn:renorm_all_coupling}) shows that the anomalous
dimension is  given by $\gamma(\tg) =1$.
Therefore two susceptibility exponents are related by
$ \gamma_0 + \gamma_1 = 2$.
\par

\section{Reparametrization Identities}
Our key observation in using the renormalization group equation
is the ambiguity%
to identify the renormalization
group flow in the coupling constant space\cite{HIS}.
Though the above equation (\ref{eqn:renorm_all_coupling}) seems to
describe a renormalization group flow in the infinite dimensional
coupling constant space,
the direction of the flow is in fact ambiguous
because all the differential operators $(\partial/\partial\tilde{g_k})$
are not linearly independent.
Since the model is $O(N)$ invariant, we can obtain new informations
only from reparametrizations of the radial coordinate $x=\phi^2$
which become reparametrizations of a half real line.
In order to facilitate the derivation of the reparametrization
identities, it is useful to make the following reparametrization
\begin{equation}
 \phi^2 \rightarrow {\phi'}^2 = \phi^2
\left( 1+ \varepsilon {1 \over \zeta -\phi^2}\right),
 \label{eqn:reparametrization}
\end{equation}
where $\varepsilon$ is an infinitesimal parameter and $\zeta$ is a
parameter to generate all possible reparametrizations as a power series.
Substituting (\ref{eqn:reparametrization})
in (\ref{eqn:radial_int}), we obtain an identity
\begin{equation}
\left( {N \over 2}-1 \right) \VEV{{1 \over \zeta - \phi^2} }
+ \VEV{{\zeta \over (\zeta-\phi^2)^2}} -N
\VEV{{\phi^2 \over \zeta-\phi^2} V'(\phi^2)}=0,
 \label{eqn:vector_loopequation}
\end{equation}
\begin{equation}
\VEV{O}\equiv {1 \over Z_N(g)}\int d^N\phi~ O~ {\rm e}^{-NV(\phi^2)}.
\end{equation}
we shall call this identity
a loop equation since it resembles
that in the matrix model \cite{GIMO}.
It is convenient to define the expectation value $W(\zeta)$ of
a resolvent %
\be
W(\zeta)=\VEV{1 \over \zeta-\phi^2}
=\sum_{j=0}^{\infty}{\VEV{(\phi^2)^j} \over \zeta^{j+1}}
={1 \over \zeta}+\sum_{j=1}^{\infty}{2j \over \zeta^{j+1}}
 \left. {\partial F(N,g) \over \partial g_j}\right|_g
\ee
By expanding the identity (\ref{eqn:vector_loopequation}),
we can obtain various reparametrization identities which form a
representation of the Virasoro algebra \cite{HIS}.
We see that derivatives of the free energy in terms of
infinitely many coupling constants $\tilde g_k$
are related by
infinitely many reparametrization identities.
Thus one can expect that only a finite number of derivatives are
linearly independent.
The loop equation for the vector model
 (\ref{eqn:vector_loopequation}) can be solved as
\begin{equation}
W(\zeta)={2 \over 1-2\zeta V'(\zeta)}
\left[\sum_{k=1}^{\infty}{d^k (\zeta V'(\zeta)) \over d\zeta^k}
\VEV{(\zeta-\phi^2)^{k-1}}-{1 \over N}\zeta W'(\zeta)\right].
 \label{eqn:solution_resolvent}
\end{equation}
The resolvent to the leading order
in $1/N$ expansion is given by omitting the last term.

To illustrate the use of the reparametrization identities,
we shall first take the case of a single coupling constant.
Let us consider a point in the coupling constant space
$(g_1, \tg_2, \tg_3, \tg_4, \ldots) = (g_1,\tg_2,0,0, \ldots)$.
The coupling constant shifts $\delta g_k$ given by eq.(\ref{eqn:shift})
are then found to be
\be
N\delta g_k= (-1)^{k+1}(\tg_2 g_1)^k + g_1 \delta_{k,1} +\tg_2 g_1^2
\delta_{k,2}.
\ee
Inserting the shifts $\delta g_k$ to the right hand side $G$ of the
renormalization group equation (\ref{eqn:renorm_all_coupling}), we find
\ba
G&\!\!\!=&\!\!\!
 - \frac{1}{2}  + {1 \over 2}\int_{-\infty}^{-1/(\tg_2 g_1)}
d \zeta \left( W(\zeta)-{1 \over \zeta}\right)
+g_1  \left. {\partial F \over \partial g_1} \right|_{g}
+\tg_2 g_1^2 \left. {\partial F \over \partial g_2} \right|_{g} \nonu
&\!\!\!=&\!\!\!
 {1 \over 2}\int_{-\infty}^{-1/(\tg_2 g_1)}
d \zeta \left( W(\zeta)-{1 \over \zeta}\right)
-\tg_2 \left. {\partial F \over \partial \tg_2} \right|_{\tg}
\ea
By using the solution for the resolvent (\ref{eqn:solution_resolvent})
to the leading order in $1/N$, we find that the renormalization group
flow in infinitely many coupling constant space is now reduced to an
effective flow in single coupling constant $\tg_2$.
Thus we obtain a renormalization group equation with
the effective beta function $\beta^{\mbox{eff}}(\tg_2)$
and the inhomogeneous term $r(\tilde g_2)$
\begin{equation}
\left[N\frac{\partial}{\partial N}
+ 1 \right]  F(N,\tg_2) = G =
\beta^{\mbox{eff}}(\tg_2)
\frac{\partial  F(N,\tg_2)}{\partial \tg_2}
+r( \tg_2),
  \label{eqn:effective_rge}
\end{equation}
where the effective beta function and the inhomogeneous term are given
in terms of $\Delta =  4 \tg_2 + 1$ as
\begin{eqnarray}
\beta^{\mbox{eff}}(\tg_2)
&\!\!\! = &\!\!\! {1 \over 4}
\left[1-\Delta - {(1-\Delta)^2 \over 2\sqrt{\Delta}}
\log \left({1+\sqrt{\Delta} \over 1-\sqrt{\Delta}}\right)\right]
\label{eqn:effectivebeta}
\\
r(\tg_2)
&\!\!\! = &\!\!\! {1 \over 2\sqrt{\Delta}}
\left[\left({1+\sqrt{\Delta} \over 2}\right)^2
\log \left({1+\sqrt{\Delta} \over 2}\right)
-\left({1-\sqrt{\Delta} \over 2}\right)^2
\log \left({1-\sqrt{\Delta} \over 2}\right)\right] .
\label{eqn:inhomogeneousterm}
\end{eqnarray}

The effective beta function
$\beta^{\mbox{eff}}(\tilde g_2)$ exhibits a zero at
$\tg_2 = -1/4 \equiv g_{2*}$.
Furthermore, we get the susceptibility exponents
$
\gamma_1 = 3/2,
\gamma_0 = 1/2
$
from the derivative of the effective beta function
by using (\ref{eqn:vectorgamma_and_beta}) .
The fixed point and the susceptibility exponents
are in complete
agreement with the exact results for the $m=2$ critical point of
the vector model corresponding to pure gravity \cite{VM0}.
The beta function $\beta^{\mbox{eff}}(\tilde g_2)$ has also a trivial
fixed point at $\tg_2 = 0$, which is ultraviolet unstable since
$\partial \beta^{\mbox{eff}}/\partial \tilde g_2 (\tilde g_2=0)=-1<0$.

By a systematic expansion of the free energy in powers of $1/N$,
we can extract the complete information from the renormalization
group flow namely from the difference equation
and the reparametrization identities
\begin{equation}
F(N,\tg_2) = \sum_{h=0}^{\infty} N^{-h} f_h(\tg_2).
\end{equation}
We should
evaluate
the difference equation (\ref{eqn:difference_equation})
and solve
the loop equation (\ref{eqn:solution_resolvent})
both up to the order $N^{-h}$.
Then we obtain an ordinary differential equation for $f_h$
\be
  (1-h)f_h(\tg_2) -
\beta^{\mbox{eff}}(\tg_2)\frac{\partial f_0}{\partial \tg_2}(\tg_2)
= r_h(\tg_2),
\ee
where the effective beta function is common to all $h$,
but the inhomogeneous terms $r_h$ depend on $h$.
We see immediately that the general solution is given by a sum
of an arbitrary multiple of the solution of the homogeneous equation and
a particular solution of the inhomogeneous equation.
Since both the effective beta function $\beta^{\mbox{eff}}(\tg_2) $
and the inhomogeneous term $r(\tilde g_2)$ are analytic in $\tg_2$
around the fixed point $\tg_{2*}$,
the singular behavior of $f_h$
comes from the solution of the homogeneous equation.
It is important to notice that the singular term
corresponding to the continuum physics (the so-called universal term)
is specified by the beta function alone.
The normalization $a_h$ of singular terms, however,
cannot be obtained from the renormalization group equation
\ba
f_h(\tg_2) &\!\!\! = &\!\!\!
f_h(\tg_2)_{sing}+f_h(\tg_2)_{analytic} \nonu
f_h(\tg_2)_{sing} &\!\!\! = &\!\!\!
(\tg_2 - \tg_{2*})^{(1 -\gamma_1) h} a_h + \cdots.
\label{eqn:singularterm}
\ea
If we sum over the contributions from various $h$, we find that
the renormalization group equation determines the combinations of
variables appropriate to define the double scaling limit.
On the other hand,
the functional form of the singular part of the
free energy $F_{sing}$ in the scaling
variable
$N^{1/\gamma_1}(\tg_2-\tg_{2*})$ is
undetermined corresponding to the undetermined normalization factor
$a_h$ for the singular terms of each $h$
\ba
F(N, \tilde g)_{sing}
 &\!\!\! = &\!\!\!
\sum_{h=0}^{\infty} N^{-h}f_h(\tg_2)_{sing}
=\sum_{h=0}^{\infty} N^{-h}
(\tg_2 - \tg_{2*})^{2 - \gamma_0-\gamma_1 h} a_h \nonu
&\!\!\! = &\!\!\! (\tg_2 - \tg_{2*})^{2- \gamma_0}
f\bigl(N^{1/\gamma_1}(\tg_2-\tg_{2*})\bigr)_{sing}.
\ea
Let us note that the free energy is completely determined if we impose
the condition $F(N,\tg_2=0)=0$,  which follows from the definition
(\ref{eqn:freeenergy}).

We can extend the analysis to more general situations of finitely
many coupling constants, such as
the case of the two coupling constants
$g_1 = 1,~~ \tilde g_2\not=0,~~ \tilde g_3 \not= 0,~~
\tilde g_k = 0 ~ ( k \ge 4)$.
We can write down the shifts of coupling constants
$\delta g_k$ in eq.(\ref{eqn:shift}) explicitly, and
we can obtain the resolvent as a
solution of the loop equation (\ref{eqn:solution_resolvent}).
By inserting these to the right hand side $G$ of the
renormalization group equation, we
obtain effective beta functions and inhomogeneous terms.
As reported in ref.\cite{HIS}, we found a perfect agreement
with the exact result for $m=2, 3$ cases.
Moreover, we were able to explicitly draw the renormalization group
flow diagram in the space of two coupling constant space
\cite{HIS}. 
Higher multicritical points can be analyzed similarly.

%
\section {Nonlinear Renormalization Group Equation for Matrix Models}
In this last section we report briefly on our new results on matrix
models \cite{HINS}.
The partition function $Z_N(g)$ of the matrix model
is defined by an
integral over an $N\times N$ hermitian matrix $\Phi$
with a potential $V(\Phi)$
\be
Z_N(g)= \int d^{N^2} \Phi \exp \left[-N \tr V(\Phi) \right],
\qquad
V(\Phi)=\sum_{k=1}^{\infty}{g_k \over k}\Phi^{k}
\ee
The cubic interaction with a single coupling constant $g$
corresponds to $V(\Phi)={1 \over 2}\Phi^2+{g \over 3}\Phi^3$.
We can integrate over the angular variables to obtain an integral over
the eigenvalues $\lambda$ \cite{BIPZ}
\be
Z_N(g)=c_N \int \pr d\l_i \ \D_N^2 (\l) \ \exp \left[-N \su V(\l_i)
\right]
\ee
where $\D_N$ denotes the Van der Monde determinant
$\D_N(\l)=\prod_{1\leq i < j\leq N}(\l_i-\l_j)$, and
$c_N=\pi^{N(N-1)/2}/\prod_{p=1}^{N} p!$ .

In order to relate $Z_{N+1}$ to $Z_{N}$, we shall integrate
one of the eigenvalues $\lambda_{N+1}$ in $Z_{N+1}$
\ba
Z_{N+1}(g)&=& \int d^{(N+1)^2} \Phi \exp \left[-(N+1)\tr V(\Phi)\right]
\nonumber\\
&=& c_{N+1} \int \prod_{i=1}^{N+1} d\l_i \ \D_{N+1}^2 (\l) \
\exp \left[-(N+1) \sum_{i=1}^{N+1} V(\l_i) \right]
\ea
\be
Z_{N+1}(g)=c_{N+1} \int \prod_{i=1}^{N} d\l_i \ \D_{N}^2 (\l)
{\rm e}^{-(N+1) \sum_{i=1}^{N} V(\l_i)}
 \int d\lambda_{N+1} \pr \vert \lambda_{N+1} -\l_i \vert^2
{\rm e}^{-(N+1) V(\lambda_{N+1})}
\ee
The $\lambda_{N+1}$ integral can be evaluated by a saddle point
method to the leading order in $1/N$.
The saddle point $\lambda_{N+1}^s$ is determined as a function of
$\tr \Phi^n$ by the saddle point equation
\be
V'(\lambda_{N+1}^{s})={2 \over N} \tr \frac{1}{\lambda_{N+1}^{s}-\Phi} =
{2 \over N}\sum_{n=0}^{\infty}
{\tr \Phi^n \over (\lambda_{N+1}^s)^{n+1}}
\label{6}
\ee
By inserting the saddle point $\lambda_{N+1}^s$ into the partition
function, we obtain
\ba
\frac{Z_{N+1}(g)}{Z_N(g)}&=&
 \frac{c_{N+1}}{c_N} \frac{\int d^{N^2} \Phi \exp \left[-(N+1)\tr V(\Phi)
-N V(\lambda_{N+1}^s)+2\tr \log \vert \lambda_{N+1}^s-\Phi \vert +O(N^0)
\right] }{
\int d^{N^2} \Phi \exp \left[-N\tr V(\Phi)\right]} \nonumber \\
&=& \left\la \exp \left[
-\tr V(\Phi)
-N V(\lambda_{N+1}^s)+2\tr \log \vert \lambda_{N+1}^s-\Phi \vert +
\log \frac{c_{N+1}}{c_{N}}
+O(N^0) \right] \right\ra
\label{7}
\ea
Here $\la \ \ \ \ra$ represents the normalized average with
respect to the measure
$ d^{N^2} \Phi \exp \left[-N\tr V(\Phi)\right]$.
We observe from the above equation
and eq.(\ref{6}) that infinitely many numbers of operators of the form
$\tr \Phi^m \tr \Phi^n \cdots$ are induced.

The situation, however, simplifies if we appeal to the large-$N$ limit.
In this limit a multi-point function of $U(N)$-invariant operators
factorizes into a product of one-point functions
\begin{equation}
\la {\cal O}\ {\cal O}'  \cdots \ra
= \la {\cal O} \ra \la {\cal O}' \ra \cdots + O\lt \frac{1}{N^2}\rt .
\end{equation}
In favor of this factorization property, eq.(\ref{7}) reads
\be
\frac{Z_{N+1}(g)}{Z_N(g)}=
\exp \left[ -\la \tr V(\Phi) \ra - N
V(\langle \lambda_{N+1}^s \rangle)
+2\la \tr \log \vert \langle \lambda_{N+1}^s \rangle -\Phi \vert
\ra +\log \frac{c_{N+1}}{c_N} +O(N^0) \right] ,
\label{eqn:ratio_partitionfunc}
\ee
where $\langle \lambda_{N+1}^s \rangle$ is determined in terms of
$\la \tr \Phi^m \ra$
by averaging eq.(\ref{6}),
\be
V'(\langle \lambda_{N+1}^s \rangle)-2\left\la \ntr
\frac{1}{\langle \lambda_{N+1}^s \rangle-\Phi}\right\ra =0 .
\label{8}
\ee

Let us introduce the free energy which is normalized to vanish for
vanishing coupling constants
\be
F(N,g)\equiv -\frac{1}{N^2}
\log \left[\frac{Z_N(g)}{Z_N(g_1=0,g_2=1,g_k=0~(k\ge 2))}\right],
\ee
where
$Z_N(g_1=0, g_2=1, g_k=0~(k\ge 2))=2^{N/2} (\pi/N)^{N^2/2}$.
By taking the
$N\rightarrow\infty$ limit, we find the following differential
equation as a renormalization group equation
for the free energy
\be
\lt N\ddn +2 \rt F(N,g)=\left\la \ntr V(\Phi)\right\ra
+V(\langle \lambda_{N+1}^s \rangle)
-2\left\la \ntr \log \left\vert \left\langle \lambda_{N+1}^s \right\ra
-\Phi \right\vert \right\ra -\frac32 +O\lt \frac1N \rt.
\label{11}
\ee
We observe that this renormalization group equation describes a flow
in infinite dimensional coupling constant space analogous to the vector
model.
We stress that this RG equation does not involve a perturbation
with respect to coupling constants, unlike the approximation schemes
proposed in ref.\cite{BRZJ}.

Similarly to the vector model, we can find infinitely many
identities expressing the reparametrization freedom of matrix variables.
In order to derive these identities, it is most convenient to
formulate in terms of the loop equations \cite{GIMO},
\cite{FKN}.
Let us define the expectation value of the resolvent %
\be
W(\zeta)=\VEV{\tr\left({1 \over \zeta-\Phi}\right)}
\ee
By a reparametrization of matrix variables, we can obtain
the loop equation to the leading order in $1/N$
explicitly
\ba
&\ &W(\zeta)^2 -V'(\zeta) W(\zeta) + Q(\zeta; V)=0,\\
&\ & Q(\zeta; V)\equiv \sum_{k \geq 1} \frac{1}{k!} V^{(k+1)}(\zeta)
\left\la \ntr (\Phi-\zeta)^{k-1} \right\ra. \nonumber
\ea
If we consider the space of single coupling constant
for the cubic interaction
$(g_1, g_2, g_3, \cdots )=(0, 1, g_3, 0, \cdots)$,
we can relate derivatives with respect to
all the other coupling constants to that with respect to the
single coupling constant $g_3$ by means of the loop equation.

Therefore we obtain to the leading order in $1/N$
\begin{eqnarray}
\lt N\ddn +2 \rt F(N,g)
& =  &
-1 - \frac{g_3}{2} \frac{\pa F}{\pa g_3}
+V(\langle \lambda_{N+1}^s \rangle)
-2 \log \la \lambda_{N+1}^s \ra
-2 \int_{-\infty}^{\la \lambda_{N+1}^s\ra} d \zeta
\left(W(\zeta) - \frac{1}{\zeta}\right) \nonumber \\
& \equiv & G \left(g_3, {\partial F_N \over \partial g_3}\right).
\label{eqn:nonlinearrengreq}
\end{eqnarray}
Contrary to the vector model, this effective renormalization group
equation contains nonlinear terms in $\partial F_N / \partial g_3$.
\be
 G \left(g_3, {\partial F \over \partial g_3}\right)
=\sum_{n=0}^{\infty} \beta_n(g_3)
\left({\partial F \over \partial g_3}\right)^n.
\ee
This nonlinearity is the most characteristic feature of the matrix
model.

In spite of the nonlinearity, we can find the fixed points and
the critical exponents in the following way \cite{HINS}.
If we expand the coefficients $\beta_n$ in powers of the coupling
constant around the fixed point $g_{3*}$, they are analytic
\be
\beta_n(g_3)=\sum_{k=0}^{\infty}\beta_{nk} (g_3-g_{3*})^k.
\label{eqn:nonlinearbeta}
\ee
The non-analyticity of the free energy should come from solving the
differential equation (\ref{eqn:nonlinearrengreq}).
We assume that the free energy has singular and analytic terms
as in the vector model (\ref{eqn:singularterm})
\be
F(N,g_3)=F_{analytic}+F_{sing}, \qquad
F_{analytic}=\sum_{k=0}^{\infty}c_k (g_3-g_{3*})^k, \quad
F_{sing}=\sum_{k=0}^{\infty}d_k (g_3-g_{3*})^{k+\gamma}.
\ee
By comparing the power series expansion of the renormalization group
equation, we find
consistency conditions for
the above expansion to be valid.
The first four conditions read
\begin{eqnarray}
0 & = & d_0 \gamma \sum_{n=1}^{\infty} \beta_{n0} n c_1^{n-1},
\mbox{\hspace{6em}}
2c_1= \sum_{n=0}^{\infty} \beta_{n1} c_1^{n}, \nonumber \\
2d_0 & = & d_0 \gamma \left[
\sum_{n=1}^{\infty} \beta_{n1} n c_1^{n-1}
+2c_2 \sum_{n=2}^{\infty} \beta_{n0} n(n-1) c_1^{n-2} \right], \\
2c_2 & = &
\sum_{n=0}^{\infty} \beta_{n2} c_1^{n}
+2c_2 \sum_{n=1}^{\infty} \beta_{n1} n c_1^{n-1}
+2(c_2)^2 \sum_{n=2}^{\infty} \beta_{n0} n(n-1) c_1^{n-2}.  \nonumber
\end{eqnarray}
By solving these four equations
we can determine four quantities, namely
the fixed point, the coefficient $c_1$, the critical exponent $\gamma$,
and the coefficient $c_2$.
Therefore these equations are enough to determine the fixed point and
the critical exponents.
We have found that the above set of equations in fact has a solution
that agrees with the exact result \cite{HINS}
$
g_{3*}=\pm 2^{-1} \cdot 3^{-3/4}, \qquad \gamma_1=\gamma= 5 / 2.
$

Similarly we can determine all the coefficients $c_k$ $(k\ge 0)$ and
$d_k/d_0$ ($k\ge 1$) except for the overall normalization of the
singular term $d_0$.
Moreover we can show that higher genus free
energies have singular terms precisely needed for the double
scaling behavior
\begin{equation}
F(N,g_3)_{sing} = \sum_{h=0}^{\infty} N^{-2h} f_h(g_3)_{sing},
\qquad
f_h(g_3)_{sing} =
(g_3 - g_{3*})^{(1 - h) \gamma_1} d_0^h + \cdots.
\end{equation}
The singular terms are determined up to the normalization $d_0^h$
for each genus.
These features are analogous to the case of the vector model.
However, a distinctive feature of the matrix model is that the two
coefficients $c_1$ and $c_2$ are needed to determine the fixed point
$g_*$ and the critical exponent $\gamma_1$.
This novel new feature is a direct consequence of the nonlinear nature
of the renormalization group equation for the matrix model
(\ref{eqn:nonlinearbeta}).
It is intriguing to observe that these two coefficients can
be determined self-consistently without referring to all the other
coefficients $c_k$ ($k\not = 1,2$).
We can perform a similar analysis for cases with more couplings
corresponding to the higher multicritical points, and also
multi-matrix models \cite{HINS}.
%
\par
\vspace{3mm}
\begin{small}
One of the authors (N.S.) thanks the organizers of the First Pacific
Winter School for Theoretical Physics, especially Y.M. Cho for their
warm hospitality.
This work is supported in part by Grant-in-Aid for Scientific
Research (S.H.) and (No.05640334)(N.S.), and Grant-in-Aid for Scientific
Research for Priority Areas (No.05230019)(N.S.) {}from
the Ministry of Education, Science and Culture.
\end{small}
\vspace{3mm}
\end{document}